# SNAP microwave optical filters

## M. SUMETSKY

*Aston Institute of Photonic Technologies, Aston University, Birmingham B4 7ET, UK*
*Corresponding author: m.sumetsky@aston.ac.uk*

**If the originally flat bottom of a wide quantum well with multiple eigenstates is periodically modulated, its eigenvalues rearrange into denser groups separated by wider gaps. We show that this effect, if implemented in an elongated bottle microresonator (also called a SNAP microresonator) allows to design microwave photonic tunable filters with an outstanding performance.**

Design and fabrication of high-quality microwave filters is a longstanding problem which attracted scientists and engineers for several decades [1-3]. The interest to this problem is motivated by critical applications of microwave filters in modern communication technologies where their accurate transmission spectrum characteristics is highly desirable to combine with small dimensions and broadband tunability. Photonics suggests several solutions to this problem based on miniature photonic circuits [4-6]. In particular, much research work was done to design and fabricate microwave filters based on coupled ring resonators [7-9], photonic crystals [10], distributed feedback resonators [11], Mach-Zehnder interferometers [12], fiber Bragg gratings [13], frequency comb generators [14], Brillouin scattering [15] and other approaches [4-6].

In many cases, it is important to create filters which transmission amplitude has maximum flatness within the predetermined bandwidth and steeply vanishes outside it. Theoretically, filters with predetermined flatness and high rejection rate can be designed by apodization of coupled microresonator circuit with sufficiently large number of elements [16, 17]. Experimentally, the intrinsic losses and insufficient fabrication precision lead to severe noise in the transmission amplitude of such circuits growing with the number of resonant elements [18] and impracticality of devices fabricated of sufficiently large number of coupled microresonators.

Even for the negligible propagation losses, at light frequency $f$ and for the pass bandwidth $\Delta f$, microresonators with characteristic dimension $d$ should be fabricated with the precision of better than $\Delta d \sim \Delta f d/f$. For characteristic $f = 200$ THz, $\Delta f = 100$ MHz, and $d = 100$ μm, we have $\Delta d \sim 0.5$ Å, not possible to achieve by conventional modern microphotonic fabrication technologies. For this reason, coupled ring resonator and other photonic infinite impulse response filters were fabricated with the aid of microheaters enable to tune the circuit elements individually (see, e.g., [9, 11, 19] and references therein).

Increasing the microresonator Q-factor allows to create filters with better passband flatness and larger rejection rate. Indeed, at optical frequency $f$, the pass bandwidth cannot be smaller than $\Delta f_{pass} \sim f/Q$, while to arrive at sufficient flatness we have to have $Q \gg f/\Delta f_{pass}$. Thus, at $f = 200$ THz, the characteristic for microwave applications passband with $\Delta f_{pass} = 100$ MHz requires $Q \gg 2 \cdot 10^6$. That high and much higher Q-factors are possible to achieve in standing along microresonators [20, 21]. However, the problem of effective combining them into a circuit of multiple elements with the predetermined dimensions and coupling remains open.

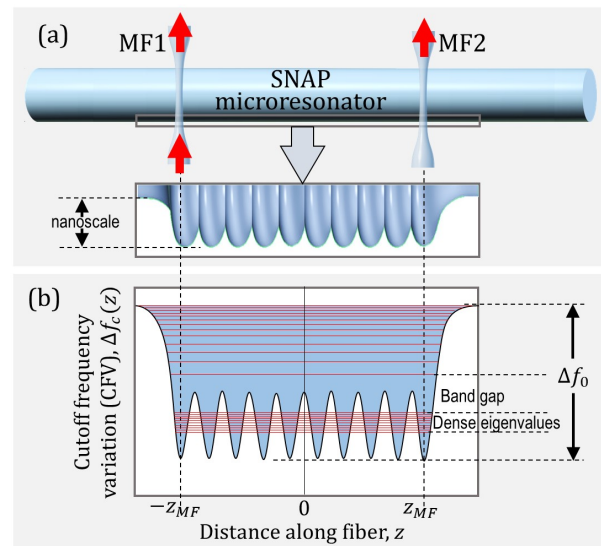

Fig. 1. (a) Illustration of a SMR with nanoscale periodically modulated ERV coupled to microfibers MF1 and MF2 and (b) the corresponding CFV. Red lines are the axial frequency eigenvalues of SMR.

Favorably, the ultraprecise fabrication precision combined with ultralow material and scattering losses required for realization of microwave photonic filters can be achieved in the surface nanoscale axial photonics (SNAP) technology [22-26] which has not yet been considered for microwave applications. In SNAP, the required microresonator circuits are fabricated in the form of coupled bottle microresonators having nanoscale effective radius variation (ERV). In Ref. [23], 30 coupled bottle microresonators were fabricated at the surface of a 19 μm radius optical fiber with better than 1 Å precision. In Ref. [25], it was shown that coupled SNAP bottle microresonators can be postprocessed with the frequency precision of better than 0.2 GHz. This fabrication precision can be further improved since in Ref. [25] it was limited by the resolution of the optical spectrum analyzer used. The intrinsic Q-factor of bottle microresonators (characterizing material and scattering losses) can be greater than the loaded Q-factor $3 \cdot 10^8$ of a bottle resonator with similar radius measured in Ref. [27]. All this suggests the SNAP technology as a promising approach for fabrication of practical microwave photonics filters and signal processing devices.

In this Letter we consider a SNAP microresonator (SMR) with periodically modulated ERV illustrated in Fig. 1(a) (which can be also considered as a system of coupled bottle microresonators) coupled to the input microfiber MF1 and output microfiber MF2. The frequency eigenvalue structure of this SMR is illustrated in Fig. 1(b). This structure has a series of dense eigenfrequencies separated from others by a gap. The experimental realization of similar SMRs with subangstrom precision was demonstrated in [22, 23], while a four-port resonant SMR device coupled to two microfibers was fabricated in Ref. [28] following the demonstration of an ultralow loss four-port microtoroid device in Ref. [29]. We show below that the periodic SMR structure illustrated in Fig. 1 can be apodized and appropriately coupled to MF1 and MF2 to perform as a bandpass filter with close to flat transmission amplitude and high rejection.

Following the SNAP theory [22], we consider whispering gallery modes (WGMs), which slowly propagate along the SMR axis $z$, and introduce the cutoff frequency $f_c(z)$ corresponding to a WGM with fixed azimuthal and axial quantum numbers, $m = m_0$ and $p = p_0$. A relatively small nonresonant transmission from MF1 to MF2 determined by WGMs with other quantum numbers, which limits the rejection rate, will be discussed later. The nanoscale ERV $\Delta r(z) = r(z) - r_0$ of a SMR with radius $r_0$ is expressed through the cutoff frequency variation (CFV) $\Delta f_c(z) = f_c(z) - f_0$ as $\Delta r(z) = r_0 \Delta f_c(z)/f_0$. Therefore, we can characterize a SMR either by ERV $\Delta r(z)$ or by CFV $\Delta f_c(z)$. We also introduce the frequency variation $\Delta f = f - f_0$ and characteristic frequency variation of our SMR $\Delta f_0$ (Fig. 1(b)). Then the one-dimensional wave equation describing the axial dependence of WGM amplitude [22] can be presented in the dimensionless form:

$$\Psi_{\zeta\zeta} + \left(\varepsilon - \nu(\zeta) + i\gamma + \Lambda\delta(\zeta - \zeta_{MF}) + \Lambda\delta(\zeta + \zeta_{MF})\right)\Psi = 0, \quad (1)$$

Here the dimensionless frequency $\varepsilon$, attenuation $\gamma$, MF-SMR coupling parameter $\Lambda$, and distance along the SMR $\zeta$ are defined as:

$$\varepsilon = \frac{\Delta f}{\Delta f_0}, \gamma = \frac{g}{\Delta f_0}, \zeta = \frac{z}{z_0}, \Lambda = z_0 D, z_0 = \frac{\pi c}{2^{3/2} n (f_0 \Delta f_0)^{1/2}}, \quad (2)$$

where $c$ is the speed of light, $n$ is the SMR refractive index (below we consider silica SMR with $n = 1.44$), $g$ is the attenuation expressed through its Q-factor as $g = f_0/Q$, and $D$ is the microfiber-SMR complex-valued coupling parameter [22], which is assumed the same for MF1 and MF2. We also assume that the SMR is symmetric with respect to its center at $z = 0$ and MF1 and MF2 are positioned symmetrically at axial coordinates $z = z_{MF}$ and $z = -z_{MF}$. Then, provided that the microfiber-SMR coupling is lossless [29, 30], the transmission amplitude $S_{12}(f, z_{MF})$ from MF1 to MF2 is determined as [22]

$$S_{12}(\Delta f, z_{MF}) = \frac{2\,\text{Im}(\Lambda)G(\varepsilon, \zeta_{MF}, -\zeta_{MF})}{\left(1 + \Lambda G(\varepsilon, \zeta_{MF}, \zeta_{MF})\right)^2 - \Lambda^2 G^2(\varepsilon, \zeta_{MF}, -\zeta_{MF})}. \quad (3)$$

Here $\zeta_{MF} = z_{MF}/z_0$, $\varepsilon = \Delta f/\Delta f_0$ and $G(\varepsilon, \zeta_1, \zeta_2)$ is the Green's function of Eq. (1). To take into account the coupling loss (which can be very small for a four-port microresonator [29]), the numerator in this equation should be reduced accordingly [22, 28]. The dimensionless form of Eqs. (1)-(3), allows us to design filters with different passbands $\Delta f_{pass}$ by rescaling. It follows from Eqs. (2) and (3) that to design a filter with passband $\sigma \Delta f_{pass}$ from a filter with passband $\Delta f_{pass}$ we have to rescale the CFV of the last filter by $\sigma$, its length by $\sigma^{-1/2}$, its Q-factor by $\sigma$, and the coupling parameter $D$ by $\sigma^{1/2}$.

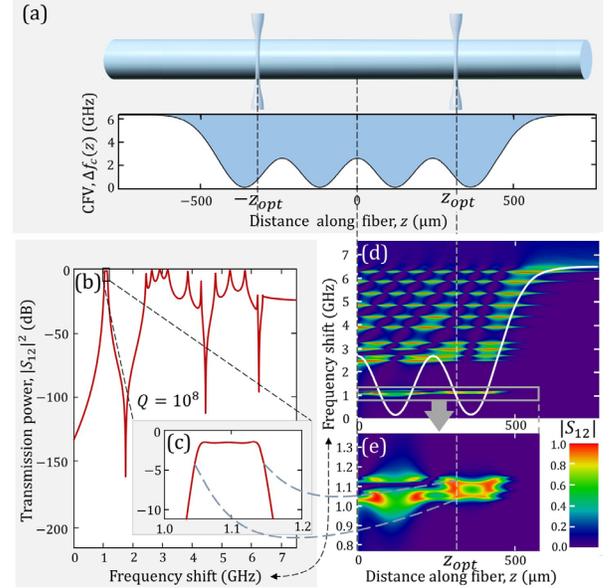

Fig. 2. (a) CFV for a SMR consisting of 4 coupled microresonators. (b) The transmission power spectrum of optimized SMR along its whole bandwidth and (c) near the passband considered. (d) The spectrogram of transmission amplitude $|S_{12}|$ calculated for the optimized coupling parameter $D$ which is magnified near the considered passband in (e).

We start with the design of a 100 MHz passband filter constructed of four coupled microresonators modeled by harmonic oscillations of CFV plotted in Fig. 2(a). Fabrication of similar SMR with subangstrom precision was demonstrated in [22, 23]. The CFV precision achieved in [25] was 0.16 GHz sufficient to introduce the CFV with characteristic 2 GHz amplitude shown in Fig. 2(a). In our modelling, we assume that the intrinsic SMR Q-factor, which determines its internal losses, is $Q = 10^8$ [27, 31] and set the frequency $f_0 = 193.4$ THz corresponding to light wavelength 1550 nm. We optimize the symmetric positions of MF1 and MF2 to arrive at the best flat transmission power $|S_{12}|^2$ within bandwidth $\Delta f_{pass} = 100$ MHz and vanishing outside it. The result of optimization is shown in Fig. 2(b) along the whole SMR spectrum and in Fig.2(c) for the spectrum in the vicinity of the passband considered. It is seen that the transmission power is quite flat within the passband and vanishes down to $-100$ dB in its vicinity (we show below that this value is prevailed by a greater value of nonresonant transmission). The optimized positions of MF1 and MF2 are $\pm z_{opt} = \pm 381$ μm, and coupling parameter is $D = 0.0015 + 0.0017i$ μm$^{-1}$. This value of $D$ is an order of magnitude less than those typically observed for the coupling of microfiber and SMR positioned in direct contact [22, 23, 32]. We suggest that this small coupling can be achieved by placing the microfiber (or a planar waveguide) several hundred nanometers away from the SMR [30]. Remarkably, we found that the displacement of MF1 and MF2 by several microns followed by optimization of $D$ does not significantly change the behavior of $|S_{12}|^2$.

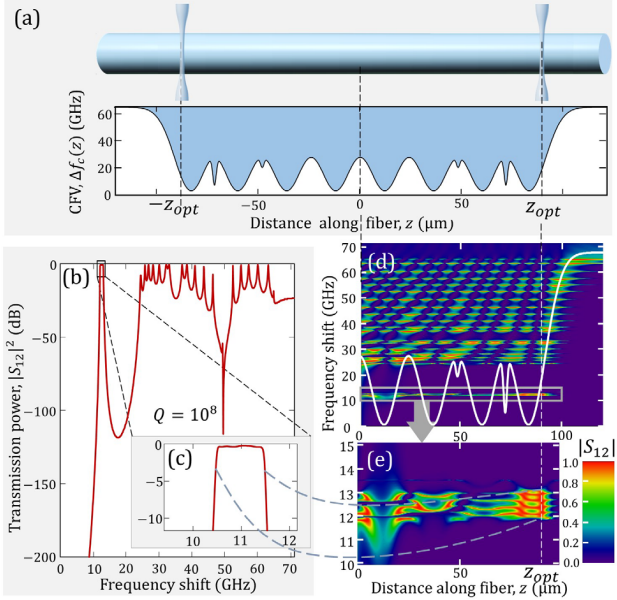

Fig. 3. (a) CFV for a SMR consisting of 8 coupled microresonators. (b) The transmission power spectrum of optimized SMR along its whole bandwidth and (c) near the passband considered. (d) The spectrogram of transmission amplitude $|S_{12}|$ calculated for the optimized coupling parameter $D$ which is magnified near the considered passband in (e).

Increasing the number of coupled microresonators and apodization of their initially periodic ERV allows to add more flexibility in designing a filter and achieve the better passband flatness and rejection rate. In the next example, we design a 1 GHz passband filter by optimization of an SMR composed of 8 coupled microresonators. Now, in addition to the optimization of positions of MF1 and MF2 made above, we apodize the SMR by narrowing the barriers between first, second, and third and, symmetrically, between sixth, seventh, and eighth microresonators. The optimized CFV of a SMR with the Q-factor $Q = 10^8$ is plotted in Fig. 3(a). While narrowing of the barriers is achieved here by straightforward cutting, we suggest that in a more advanced CFV optimization, the barrier widths can be adjusted in a way more suitable for experimental realization. It is seen from Fig. 3(a) that the length of the designed 1 GHz filter is much smaller than that of the 100 MHz filter shown in Fig. 2 and its CFV is much greater. Fig. 3(b) shows the spectrum of the transmission power $|S_{12}|^2$ along the full SMR bandwidth, which was obtained by optimization of the transmission power along the passband magnified in Fig. 3(c). The determined optimized positions of MF1 and MF2 are $\pm z_{opt} = \pm 90$ μm and the coupling parameter is $D = 0.0145 + 0.013i$ μm$^{-1}$.

The considered models and rescaling relations indicated above allow to design SMRs having CFVs and transmission spectra with other passbands. For example, to design a SMR with 500 MHz passband from the 100 MHz passband SMR with $Q = 10^8$ described above (Fig. 2), we have to magnify the frequency values along the horizontal axis in Figs. 2(b) and (c) by five. The CVW of this SMR is obtained by dividing the values of distance along the horizontal axis in Fig. 2(a) by $5^{1/2}$ (i.e., this SMR is $5^{1/2}$ times shorter) and multiplying the CFV values along the vertical axis by 5. The Q-factor of this SMR is five times smaller, $Q = 2 \cdot 10^7$.

The calculated very large transmission sideband rejection (down to -100 dB in both cases shown in Fig. 2(b) and 3(b)) is violated by non-resonant transmission of light from MF1 to MF2 not taken into account by Eq. (1) which describes the contribution of WGMs having a single azimuthal and radial quantum number, $m = m_0$ and $p = p_0$, only. To estimate the contribution of WGMs with nonresonant quantum numbers, we assume that all WGMs with radial quantum numbers greater than $p = 0$ vanish (like, e.g., in a capillary SMR with sufficiently narrow walls [33]). We introduce the separation of the cutoff frequencies $\Delta f_{az}$ along the azimuthal quantum number $m$ in the vicinity $|\Delta m| \ll m_0$ of $m_0$, where $\Delta m = m_0 - m$. Then $\Delta f_{az}$ is expressed through the SMR radius $r_0$ as $\Delta f_{az} = c(2\pi n r_0)^{-1}$. For fiber radii and CFVs of our concern, $r_0 \lesssim 1$ mm and $\Delta f_0 \lesssim 10$ GHz, we have $\Delta f_0 \ll \Delta f_{az}$. Under these assumptions, the contribution to the nonresonant transmission $S_{12}^{(nr)}$ of azimuthal modes with $\Delta m < 0$ is negligible, while the contribution of $M$ WGMs with $\Delta m > 0$ is [34]:

$$S_{12}^{(nr)} = S_0^{(nr)} \Xi, \quad \Xi = \sum_{\Delta m=1}^{M} \frac{1}{\Delta m^{1/2}} \exp\left[2i\beta_{az} z_{opt} \left(\Delta m + \frac{\Delta f}{\Delta f_{az}}\right)^{1/2}\right], \quad (4)$$

$$S_0^{(nr)} = \frac{\text{Im}(D)}{\beta_{az}}, \quad \beta_{az} = \frac{2^{2/3} \pi n}{c}(f_0 \Delta f_{az})^{1/2}, \quad \Delta f_{az} = \frac{c}{2\pi n r_0}.$$

Factor $S_0^{(nr)}$ in this equation determines the characteristic value of non-resonance transmission, while the sum over $\Delta m$ rapidly oscillates as a function of SMR radius $r_0$. For the 100 MHz passband SMR considered above (Fig. 2) with radius $r_0 = 20$ μm we have $20 \log(S_0^{(nr)}) \cong -53$ dB. For the 1 GHz filter with the same radius (Fig. 3), we have $20 \log(S_0^{(nr)}) \cong -35$ dB. These values are much greater than the rejection rates calculated above in the resonance approximation. Therefore, they determine the sideband rejection rate of the designed filters. The increase of the rejection rate with the reduction of microfiber-SMR coupling determined by $\text{Im}(D)$ correlates with experimental observations for a ring microresonator [35]. Numerical modelling based on Eq. (4) shows that these sideband transmission values can be reduced by $\sim 10$ dB by optimization of the fiber radius $r_0$.

Remarkably, connection of SMR filters in series allows to significantly reduce the non-resonant transmission in the rejection region and achieve the passband tunability. As an example, we consider a device consisting of two SMR 1 GHz passband filters designed above connected in series as illustrated in Fig. 4(a). The transmission amplitude of this filter is found as $S_{12}^{(2)} = S_{12}(\Delta f - \Delta f_{sh1}, z_{MF}) S_{12}(\Delta f - \Delta f_{sh2}, z_{MF})$. Here $\Delta f_{sh1,2}$ are the shifts of the cutoff frequencies of SMRs which can be tuned by adjacent thermal heaters shown in Fig. 4(a). This allows us to tune both the central frequency and the passband width of the filter. Figs. 4(b) and (c) show the spectra of transmission power of this filter for $\Delta f_{sh1} = 0$ and $\Delta f_{sh2} = 0$ (1 GHz passband, red curves), $\Delta f_{sh2} = 0.5$ GHz (0.5 GHz passband, blue curves), and $\Delta f_{sh2} = 0.8$ GHz (0.2 GHz passband, green curves). The dashed horizontal line in Fig. 4(b) shows the estimated rejection rate $\sim 70$ dB limited by the non-resonant transmission with the amplitude $S_0^{(nr)}$ defined by Eq. 4.

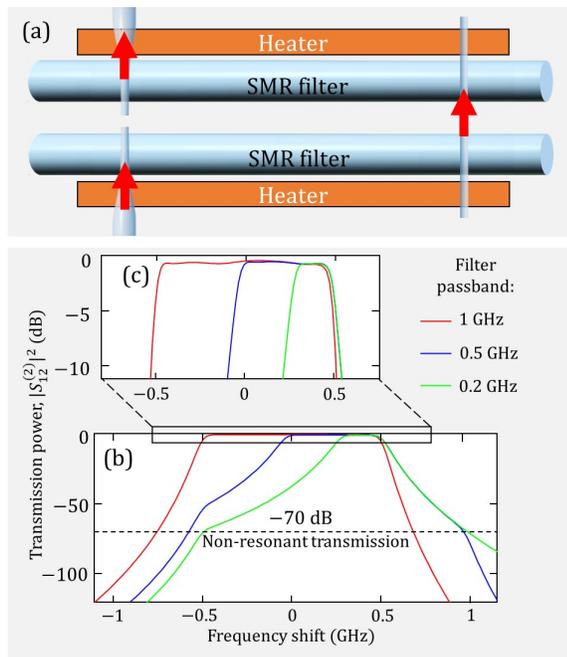

Fig. 4. (a) A tunable filter comprised of two 1 GHz filters connected in series. (b) Transmission power spectra of this filter tuned to 1 GHz, 0.5 GHz, and 0.2 GHz passbands. Dashed line shows the estimated rejection rate limited by the non-resonant transmission. (c) The same spectra magnified near the passbands.

We suggest that optimization of a SMR profile with more flexible CFV parameters, as well as SMRs connected in series, will lead to filter designs with superior transmission characteristics. Experimentally, optimization of the MF1 and MF2 positions and their coupling to SMR can be performed by their translation along the microfiber and SMR directions [24] and by tuning the microfiber-SMR gap [30]. Finally, the design and experimental realization of SMR devices for more general microwave photonics spectral shaping [36] may be of special interest.

**Funding.** Engineering and Physical Sciences Research Council, (EP/P006183/1).

**Disclosures**. The author declares no conflicts of interest.

### References

1. L. Zhu, S. Sun, and R. Li, *Microwave bandpass filters for wideband communications*, (John Wiley & Sons, Inc., 2012).
2. R. J. Cameron, C. M. Kudsia, and R. R. Mansour, *Microwave filters for communication systems: fundamentals, design, and applications*, 2nd Edition (John Wiley & Sons, Inc., 2018).
3. H. Islam, S. Das, T. Bose, and T. Ali, "Diode based reconfigurable microwave filters for cognitive radio applications: a review," IEEE Access **8**, 185429 (2020).
4. D. Marpaung, J. Yao, and J. Capmany, "Integrated microwave photonics," Nature Photon. **13**, 80 (2019).
5. Y. Liu, A. Choudhary, D. Marpaung, and B. J. Eggleton, "Integrated microwave photonic filters," Adv. Opt. Photon. **12**, 485 (2020).
6. G. Serafino, S. Maresca, C. Porzi, F. Scotti, P. Ghelfi, and A. Bogoni, "Microwave photonics for remote sensing: from basic concepts to high-level functionalities," J. Lightwave Technol. **38**, 5339 (2020).
7. B. E. Little, S. T. Chu, H. A. Haus, J. Foresi, and J.-P. Laine, "Microring Resonator Channel Dropping Filters," J. Lightwave Technol. **15**, 998 (1997).
8. W. Bogaerts, P. De Heyn, T. Van Vaerenbergh, K. De Vos, S. K. Selvaraja, T. Claes, P. Dumon, P. Bienstman, D. Van Thourhout, and R. Baets, "Silicon microring resonators," Laser Photonics Rev. **6**, 47 (2012).
9. T. A. Huffman, G. M. Brodnik, C. Pinho, S. Gundavarapu, D. Baney, D. J. Blumenthal, "Integrated resonators in an ultralow loss Si$_3$N$_4$/SiO$_2$ platform for multifunction applications," IEEE J. Sel. Top. Quantum Electron. **24**, 5900209 (2018).
10. J. Sancho, J. Bourderionnet, J. Lloret, S. Combrié, I. Gasulla, S. Xavier, S. Sales, P. Colman, G. Lehoucq, D. Dolfi, J. Capmany and A. De Rossi, "Integrable microwave filter based on a photonic crystal delay line," Nature Commun. **3**, 1075 (2012).
11. C. Porzi, G. J. Sharp, M. Sorel, and A. Bogoni, "Silicon photonics high-order distributed feedback resonators filters," IEEE J. Quant. Electron. **56**, 6500109 (2020).
12. E. J. Norberg, R. S. Guzzon, J. S. Parker, L. A. Johansson, and L. A. Coldren, "Programmable Photonic Microwave Filters Monolithically Integrated in InP–InGaAsP," J. Lightwave Technol. **29**, 1611 (2011).
13. C. Wang and J. Yao, "Fiber Bragg gratings for microwave photonics subsystems," Opt. Express **21**, 22868 (2013).
14. M. Tan, X. Xu, J. Wu, R. Morandotti, A. Mitchell and D. J. Moss, "Photonic RF and microwave filters based on 49 GHz and 200 GHz Kerr microcombs," Opt. Commun. **465**, 125563 (2020).
15. Y. Xie, A. Choudhary, Y. Liu, D. Marpaung, K. Vu, P. Ma, D.-Y. Choi, S. Madden, and B. J. Eggleton, "System-Level Performance of Chip-Based Brillouin Microwave Photonic Bandpass Filters," J. Lightwave Technol. **37**, 5246 (2019).
16. M. Sumetsky and B. J. Eggleton, "Modeling and optimization of complex photonic resonant cavity circuits," Opt. Express **11**, 381 (2003).
17. H.-C. Liu and A. Yariv, "Synthesis of high-order bandpass filters based on coupled-resonator optical waveguides (CROWs)," Opt. Express **19**, 17653 (2011).
18. W. Bogaerts, M. Fiers, and P. Dumon, "Design Challenges in Silicon Photonics," IEEE J. Sel. Top. Quant. Electron. **20**, 8202008 (2014).
19. A. A. Savchenkov, V. S. Ilchenko, A. B. Matsko, and L. Maleki, "High-Order Tunable Filters Based on a Chain of Coupled Crystalline Whispering Gallery-Mode Resonators," Photon. Technol. Lett. **17**, 136 (2005).
20. D. V. Strekalov, C. Marquardt, A. B. Matsko, H. G. L. Schwefel, and G. Leuchs, "Nonlinear and quantum optics with whispering gallery resonators," J. Opt. **18**, 123002 (2016).
21. L. Wu, H. Wang, Q. Yang, Q. Ji, B. Shen, C. Bao, M. Gao, and K. Vahala, "Greater than one billion Q factor for on-chip microresonators," Opt. Lett. **45**, 5129 (2020).
22. M. Sumetsky, "Theory of SNAP devices: basic equations and comparison with the experiment," Opt. Express **20**, 22537 (2012).
23. M. Sumetsky and Y. Dulashko, "SNAP: Fabrication of long coupled microresonator chains with sub-angstrom precision," Opt. Express **20**, 27896 (2012).
24. M. Sumetsky, "Delay of Light in an Optical Bottle Resonator with Nanoscale Radius Variation: Dispersionless, Broadband, and Low Loss," Phys. Rev. Lett. **111**, 163901 (2013).
25. N. A. Toropov and M. Sumetsky, "Permanent matching of coupled optical bottle resonators with better than 0.16 GHz precision," Opt. Lett. **41**, 2278 (2016).
26. Q. Yu, Z. Zhang, and X. Shu, "SNAP structures fabricated by profile design of in-fiber inscribed regions with a femtosecond laser," Opt. Lett. **46**, 1005 (2021).
27. M. Pöllinger, D. O'Shea, F. Warken, and A. Rauschenbeutel, "Ultrahigh-Q tunable whispering-gallery-mode microresonator," Phys. Rev. Lett. **103**, 053901 (2009).
28. M. Crespo-Ballesteros, Y. Yang, N. Toropov, and M. Sumetsky, "Four-port SNAP microresonator device," Opt. Lett. **44**, 3498 (2019).

29. H. Rokhsari and K. J. Vahala, "Ultralow loss, high Q, four port resonant couplers for quantum optics and photonics," Phys. Rev. Lett. **92**, 253905 (2004).
30. S. M. Spillane, T. J. Kippenberg, O. J. Painter, and K. J. Vahala, "Ideality in a fiber-taper-coupled microresonator system for application to cavity quantum electrodynamics," Phys. Rev. Lett. **91**, 043902 (2003).
31. M. L. Gorodetsky, A. D. Pryamikov, and V. S. Ilchenko, "Rayleigh scattering in high-Q microspheres," J. Opt. Soc. Am. **B 17**, 1051 (2000).
32. D. L. P. Vitullo, S. Zaki, D. E. Jones, M. Sumetsky, and M. Brodsky, "Coupling between waveguides and microresonators: the local approach," Opt. Express **28**, 25908 (2020).
33. G. Gardosi, B. J. Mangan, G. S. Puc, and M. Sumetsky, "Photonic microresonators created by slow optical cooking," ACS Photonics **8**, 436 (2021).
34. Eq. (4) is found by generalization of Eq. (A1.7) of Ref. [22], by adding terms determined by Green's functions with $m \neq m_0$ from Eq. (A1.3) of Ref. [22]. The expression for these Green's functions are given by Eq. (17) of the same reference. We also assume that the coupling parameter $D$ for the WGMs with $m$ close to $m_0$ is the same.
35. H. Qiu, F. Zhou, J. Qie, Y. Yao, X. Hu, Y. Zhang, X. Xiao, Y. Yu, J. Dong, and X. Zhang, "A continuously tunable sub-gigahertz microwave photonic bandpass filter based on an ultra-high-Q silicon microring resonator," J. Lightwave Technol. **36**, 4312 (2018).
36. O. Daulay, G. Liu, X. Guo, M. Eijkel, and D. Marpaung, "A Tutorial on Integrated Microwave Photonic Spectral Shaping," J. Lightwave Technol. **39**, 700 (2021).